\documentclass[12pt,aps,prd,preprint,tightenlines,superscriptaddress,
amsmath,amssymb,nofootinbib]{revtex4-1} 

\RequirePackage[colorlinks=true
,urlcolor=blue
,anchorcolor=blue
,citecolor=blue
,filecolor=blue
,linkcolor=blue
,menucolor=blue
,pagecolor=blue
,linktocpage=true
,pdfproducer=medialab
,pdfa=true
]{hyperref}

\allowdisplaybreaks

\usepackage{amsmath,amssymb,amsthm,amsfonts}
\usepackage{graphicx}
\usepackage{subfigure}
\usepackage{dcolumn}	
\usepackage{hyperref}      	
\usepackage{bm}		
\usepackage{epsfig}
\usepackage{epstopdf}
\usepackage{setspace}
\usepackage[usenames, dvipsnames]{color}
\usepackage{slashed}
\usepackage{comment}
\usepackage{enumitem}
\usepackage{siunitx}
\usepackage{multirow}
\usepackage{datetime}
\usepackage{enumitem}
\setlist[description]{leftmargin=0.4cm}

\newcommand{\PRE}[1]{{#1}} 

\newcommand{\be}{\begin{equation}\begin{aligned}}
\newcommand{\ee}{\end{aligned}\end{equation}}
\newcommand{\beq}{\begin{equation}}
\newcommand{\eeq}{\end{equation}}
\newcommand{\beqa}{\begin{eqnarray}}
\newcommand{\eeqa}{\end{eqnarray}}

\newcommand{\ifb}{\text{fb}^{-1}}
\newcommand{\iab}{\text{ab}^{-1}}

\newcommand{\mev}{\text{MeV}}
\newcommand{\gev}{\text{GeV}}
\newcommand{\tev}{\text{TeV}}

\newcommand{\micm}{\mu\text{m}}

\newcommand{\cm}{\text{cm}}
\newcommand{\m}{\text{m}}

\newcommand{\mrad}{\text{mrad}}
\newcommand{\murad}{\mu\text{rad}}

\renewcommand{\eqref}[1]{Eq.~(\ref{eq:#1})}

\newcommand{\figref}[1]{Fig.~\ref{fig:#1}}

\newcommand{\tableref}[1]{Table~\ref{table:#1}}

\RequirePackage[normalem]{ulem}

\begin{document}

\preprint{Input to the European Particle Physics Strategy \hspace*{2in} UCI-TR-2019-01}
\preprint{Update 2018-2020, Submitted 18 December 2018 \hfill KYUSHU-RCAPP-2018-08}

\title{
{\Large FASER: ForwArd Search ExpeRiment at the LHC}
\PRE{\vspace*{0.5in} \\
FASER Collaboration}
}

\begin{figure*}[h]
\centering
\vspace*{.4in}
\includegraphics[width=0.6\textwidth]{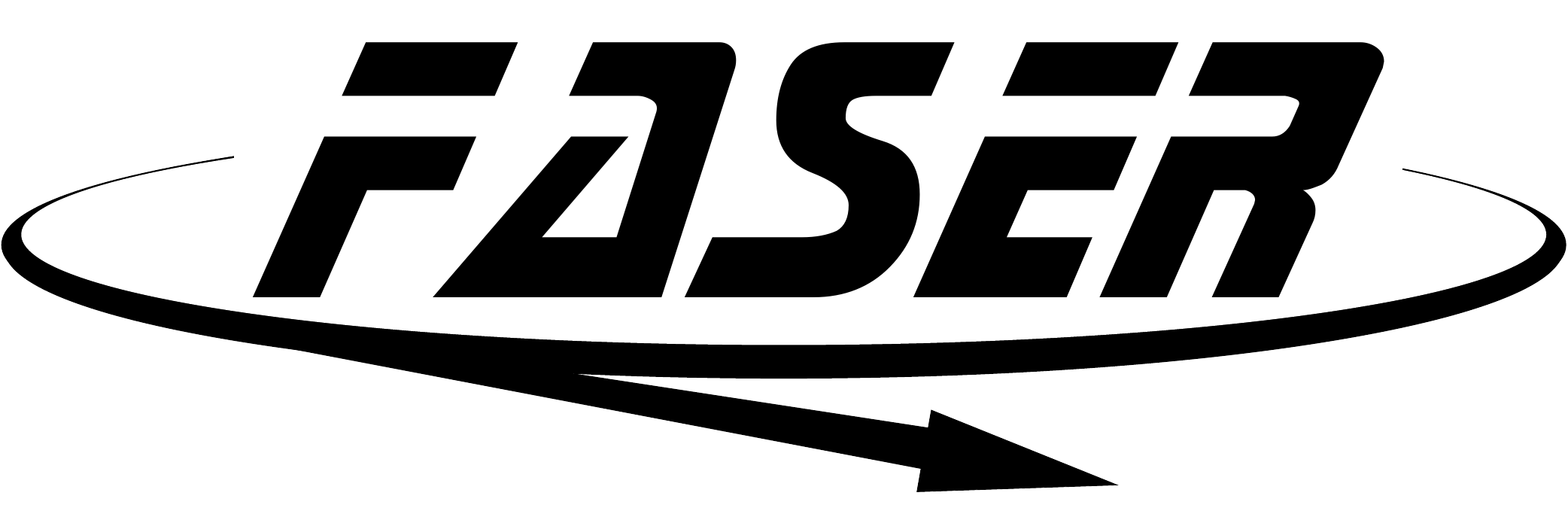}
\vspace*{.2in}
\end{figure*}

\author{Akitaka Ariga}

\author{Tomoko Ariga}

\author{Jamie Boyd}
\email[Contact email: ]{Jamie.Boyd@cern.ch}

\author{Franck Cadoux}

\author{David~W.~Casper}

\author{Yannick Favre}

\author{Jonathan~L.~Feng}
\email[Contact email: ]{jlf@uci.edu}

\author{Didier Ferrere}

\author{Iftah Galon}

\author{Sergio Gonzalez-Sevilla}

\author{Shih-Chieh Hsu}

\author{Giuseppe Iacobucci}

\author{Enrique Kajomovitz}

\author{Felix Kling}

\author{Susanne Kuehn}

\author{Lorne Levinson}

\author{Hidetoshi Otono}

\author{Brian Petersen}

\author{Osamu Sato}

\author{Matthias Schott}

\author{Anna Sfyrla}

\author{Jordan Smolinsky}

\author{Aaron~M.~Soffa}

\author{Yosuke Takubo}

\author{Eric Torrence}

\author{Sebastian Trojanowski}

\author{Gang Zhang\PRE{\vspace*{.5in}}}

\begin{abstract}
\PRE{\vspace*{0.2in}}
FASER, the ForwArd Search ExpeRiment, is a proposed experiment dedicated to searching for light, extremely weakly-interacting particles at the LHC.  Such particles may be produced in the LHC's high-energy collisions in large numbers in the far-forward region and then travel long distances through concrete and rock without interacting.  They may then decay to visible particles in FASER, which is placed 480 m downstream of the ATLAS interaction point.  In this work, we describe the FASER program.  In its first stage, FASER is an extremely compact and inexpensive detector, sensitive to decays in a cylindrical region of radius $R = 10~\cm$ and length $L = 1.5~\m$. FASER is planned to be constructed and installed in Long Shutdown 2 and will collect data during Run 3 of the 14 TeV LHC from 2021-23.  If FASER is successful, FASER 2, a much larger successor with roughly $R \sim 1~\m$ and $L \sim 5~\m$, could be constructed in Long Shutdown 3 and collect data during the HL-LHC era from 2026-35.  FASER and FASER 2 have the potential to discover dark photons, dark Higgs bosons, heavy neutral leptons, axion-like particles, and many other long-lived particles, as well as provide new information about neutrinos, with potentially far-ranging implications for particle physics and cosmology.  We describe the current status, anticipated challenges, and discovery prospects of the FASER program.  
\end{abstract}


\maketitle


\clearpage

\section{Scientific Context and Objectives}
\label{sec:introduction}

For decades, a focus of energy-frontier particle colliders, such as the LHC, has been searches for new particles with TeV-scale masses and ${\cal O}(1)$ couplings.  The common lore was to target large transverse momentum ($p_T$) signatures that emerge in the roughly isotropic decays of such particles.  There is, however, a complementary class of viable new particles that are much lighter, with masses in the MeV to GeV range, and much more weakly coupled to the standard model (SM)~\cite{Battaglieri:2017aum}. In recent years, these particles have attracted growing interest, in part because they can yield dark matter with the correct relic density~\cite{Boehm:2003hm,Feng:2008ya} and may resolve discrepancies between low-energy experiments and theoretical predictions~\cite{Bennett:2006fi, Pohl:2010zza, Krasznahorkay:2015iga}. Perhaps most importantly, they can be discovered at a wide variety of experiments, reinvigorating efforts to find creative ways to search for new particles.

If new particles are light and very weakly coupled, the focus at the LHC on particle searches at high $p_T$ may be completely misguided.  In contrast to TeV-scale particles, which are produced more or less isotropically, light particles with masses in the MeV to GeV range are dominantly produced at low $p_T \sim 100~\mev - \gev$.  In addition, because the new particles are extremely weakly coupled, very large SM event rates are required to discover the rare new physics events.  These rates are available, not at high $p_T$, but at low $p_T$: at the 13 TeV LHC, the total inelastic $pp$ scattering cross section is $\sigma_{\text{inel}}(13~\text{TeV}) \approx 75~\text{mb}$~\cite{Aaboud:2016mmw, VanHaevermaet:2016gnh}, with most of it in the far-forward direction. In upcoming runs at 14 TeV, where the inelastic cross section is very similar, we expect
\be
N_{\text{inel}} \approx 1.1 \times 10^{16} \ (2.2 \times 10^{17})
\label{eq:ppcollisions}
\ee
inelastic $pp$ scattering events for an integrated luminosity of $150~\text{fb}^{-1}$ at LHC Run 3 ($3~\text{ab}^{-1}$ at the HL-LHC). This, in turn, implies extraordinary production rates for low-mass particles; for example, for the HL-LHC, the number of mesons produced in each hemisphere is
\be
N_{\pi^0} \approx 4.6 \times 10^{18}, \ \ N_{\eta} \approx 5.0 \times 10^{17} , 
\ \ N_D \approx 2.2 \times 10^{16} , \ \ \text{and} \ \ N_B \approx 1.4 \times 10^{15} \ . \ 
\ee
Note that the number of $B$ mesons exceeds the number available at fixed target experiments, where the rate is highly suppressed by lower center-of-mass energies. 

We therefore find that even extremely weakly-coupled new particles may be produced in sufficient numbers in the far-forward region.  Given their weak coupling to the SM, such particles are typically long-lived and travel a macroscopic distance before decaying back into SM particles.  Moreover, such particles may be highly collimated.  For example, new particles that are produced in pion or $B$ meson decays are typically produced within angles of $\theta \sim \Lambda_{\text{QCD}} / E$ or $m_B / E$ of the beam collision axis, where $E$ is the energy of the particle.  For $E \sim \text{TeV}$, this implies that even $\sim 500~\m$ downstream, such particles have only spread out $\sim 10~\cm - 1~\m$ in the transverse plane.  A small and inexpensive detector placed in the far-forward region may therefore be capable of extremely sensitive searches. 

The FASER program is specifically designed to take advantage of this opportunity.  Ideal locations for FASER exist in TI12 and TI18, existing and unused side tunnels that are 480 m downstream from the ATLAS interaction point (IP). In its first stage, FASER is an extremely small and inexpensive detector,  with a decay volume of only $0.047~\m^3$. FASER is planned to be constructed and installed in TI12 in Long Shutdown 2 (LS2) from 2019-20 in time to collect data in Run 3 from 2021-23. A larger successor experiment, FASER 2, with a decay volume of very roughly $\sim 10~\m^3$, is envisioned to be constructed and installed in LS3 from 2024-26 in time to take data during the HL-LHC era from 2026-35. Despite their relatively small size, FASER and FASER 2 will complement the LHC's existing physics program, with remarkable sensitivity to dark photons, axion-like particles, and other proposed particles. In the following sections, we discuss FASER's location, layout, and discovery potential.  Additional details may be found in FASER's Letter of Intent~\cite{Ariga:2018zuc} and Technical Proposal~\cite{Ariga:2018pin}.  In the Appendix, an Addendum to this document contains information about the interested community, anticipated construction and operating costs, and computing requirements.

\section{Location}
\label{sec:location}

The side tunnels TI12 and TI18 are nearly ideal locations for FASER~\cite{Feng:2017uoz}.  These side tunnels were formerly used to connect the SPS to the LEP (now LHC) tunnel, but they are currently unused. The LHC beam collision axis intersects TI12 and TI18 at a distance of 480 m to the west and east of the ATLAS IP, respectively.  Estimates based on detailed simulations using FLUKA~\cite{Ferrari:2005zk,Bohlen:2014buj} by CERN's Sources, Targets, and Interaction (STI) group~\cite{FLUKAstudy}, combined with {\it in situ} measurements using emulsion detectors, have now confirmed a low rate of high-energy SM particles in these locations. Additionally, the FLUKA results combined with radiation monitor measurements have confirmed low radiation levels in these tunnels.  These locations, then, provide extremely low background environments for FASER to search for LLPs that are produced at or close to the IP, propagate in the forward direction close to the beam collision axis, and decay visibly within FASER's decay volume.  

\begin{figure}[tbp]
\centering
\includegraphics[width=0.59\textwidth]{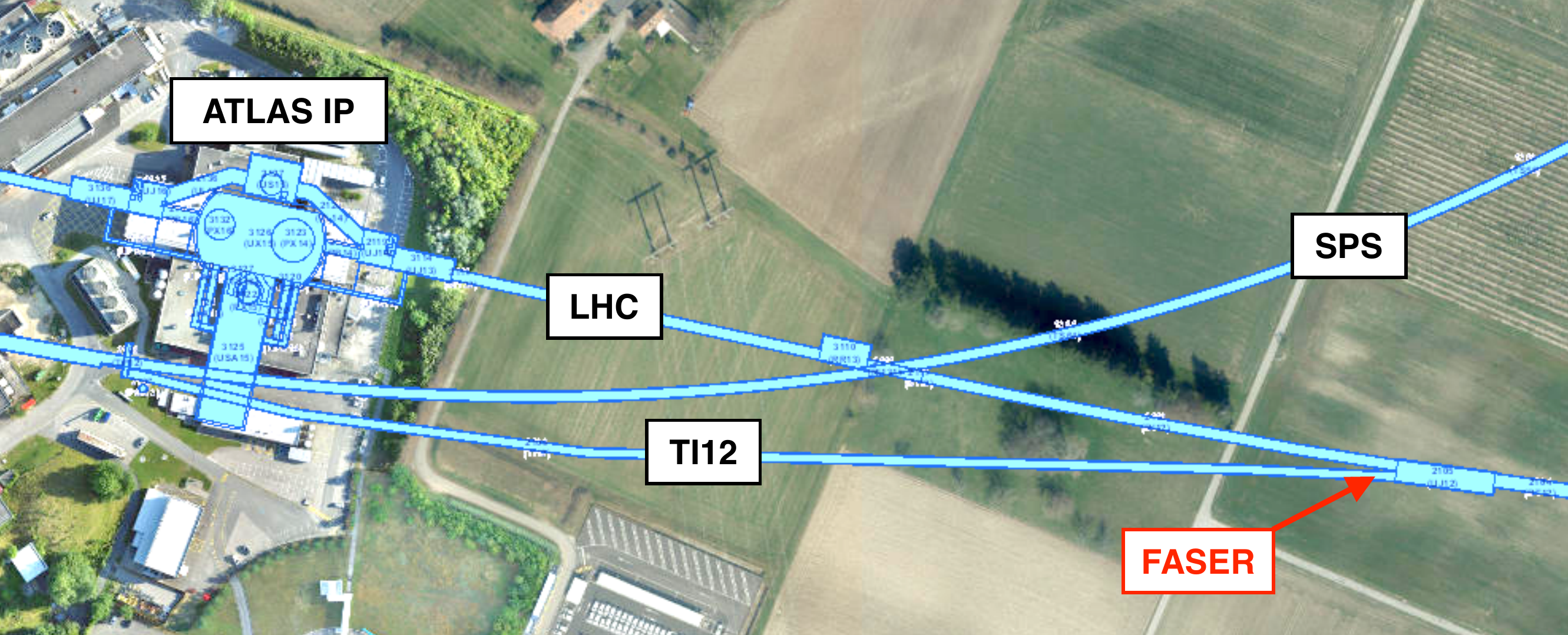} \hfill
\includegraphics[width=0.394\textwidth]{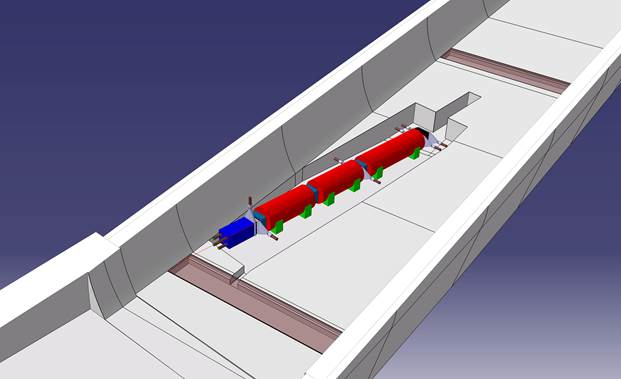}
\caption{
{\bf Left panel}: The arrow points to FASER's location in service tunnel TI12, roughly 480 m east of the ATLAS IP. Credit: CERN Geographical Information System.  {\bf Right panel}: View of FASER in tunnel TI12.  The trench lowers the floor by 45 cm at the front of FASER to allow FASER to be centered on the beam collision axis. Credit: CERN Site Management and Buildings Department. } 
\label{fig:maps}
\end{figure}

FASER is currently planned for installation in TI12.  This location is shown in \figref{maps}, and is roughly 480 m east of the ATLAS IP.  The beam collision axis passes along the floor of TI12, with its exact location depending on the beam crossing angle at ATLAS.  TI12 slopes upward when leaving the LHC tunnel to connect to the shallower SPS tunnel.  To place FASER along the beam collision axis, the ground of TI12 must be lowered roughly 45 cm at the front of FASER, where particles from the ATLAS IP enter.

\begin{figure}[tbhp]
\centering
\includegraphics[width=0.98\textwidth]{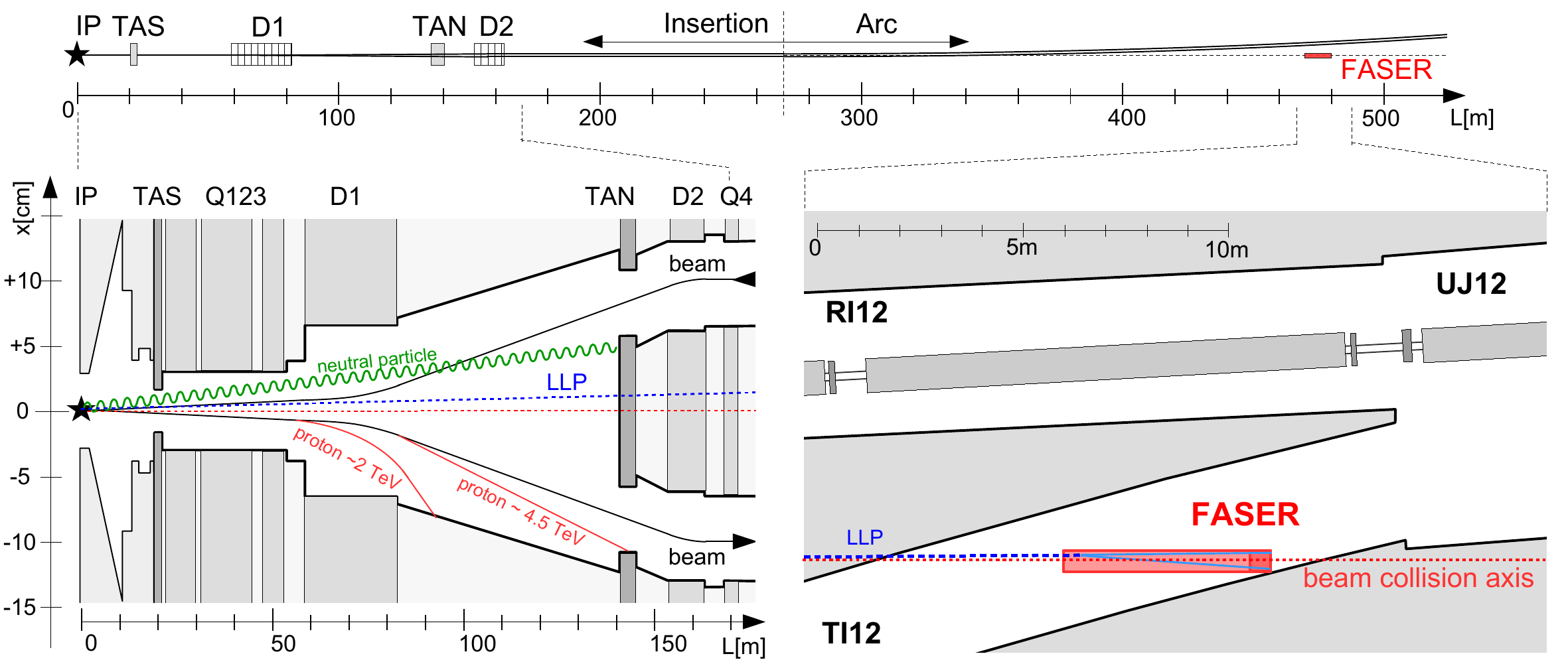} 
\caption{Schematic view of the far-forward region downstream of ATLAS and various particle trajectories. {\bf Upper panel}: FASER is located $480~\m$ downstream of ATLAS along the beam collision axis (dotted line) after the main LHC tunnel curves away.  {\bf Lower left panel}: High-energy particles produced at the IP in the far-forward direction.  Charged particles are deflected by LHC magnets, and neutral hadrons are absorbed by either the TAS or TAN, but LLPs pass through the LHC infrastructure without interacting. Note the extreme difference in horizontal and vertical scales.  {\bf Lower right panel}: LLPs may then travel $\sim 480~\m$ further downstream and decay within FASER in TI12.} 
\label{fig:infrastructure}
\end{figure}

A schematic view of the far-forward region downstream of ATLAS is given in \figref{infrastructure}.  From the ATLAS IP, the LHC beam passes through a 270 m-long straight ``insertion,'' and then enters an ``arc'' and bends.  Far-forward charged particles are bent by the beam optics, and neutral hadrons are typically stopped in the TAS or TAN absorbers, which are designed to protect the magnets.  To travel from the IP to FASER, particles must pass through roughly 10 m of concrete and 90 m of rock.  In the SM, only muons and neutrinos can reach FASER from the IP.   On the other hand, LLPs produced at or near the IP easily pass through all of the natural and man-made material without interacting and then can decay in FASER.

\section{Detector Layout and Components}

At the LHC, light particles are typically produced with a characteristic transverse momentum comparable to their mass $p_T\sim m$. Consequently, LLPs that are produced within FASER's angular acceptance, $\theta \sim p_T/E\le 1~\mrad$, tend to have very high energies $\sim \tev$.  The characteristic signal events at FASER are, then,
\begin{equation}
  p p  \to \text{LLP} +X, \quad  \text{LLP travels} \ \sim 480~\text{m}, \quad \text{LLP} \to e^+ e^- , \mu^+ \mu^- , \pi^+ \pi^-, \gamma \gamma, \ldots ,
\end{equation}
where the LLP decay products have $\sim \tev$ energies. The target signals at FASER are therefore striking: two oppositely charged tracks or two photons with $\sim \tev$ energies that emanate from a common vertex inside the detector and have a combined momentum that points back through 100 m of concrete and rock to the IP. 

The decay products of such light and highly boosted particles are extremely collimated, with a typical opening angle $\theta \sim m/E$. For example, for an LLP with mass $m \sim 100~\mev$ and energy $E \sim 1~\tev$, the typical opening angle is $\theta \sim m/E \sim 100~\murad$, implying a separation of only $\sim 100~\micm$ after traveling through $1~\m$ in the detector.  To resolve the two charged tracks produced by a decaying LLP, FASER must include a magnetic field to split the oppositely-charged tracks.  

\begin{figure}[tbp]
\centering
\includegraphics[width=0.87\textwidth]{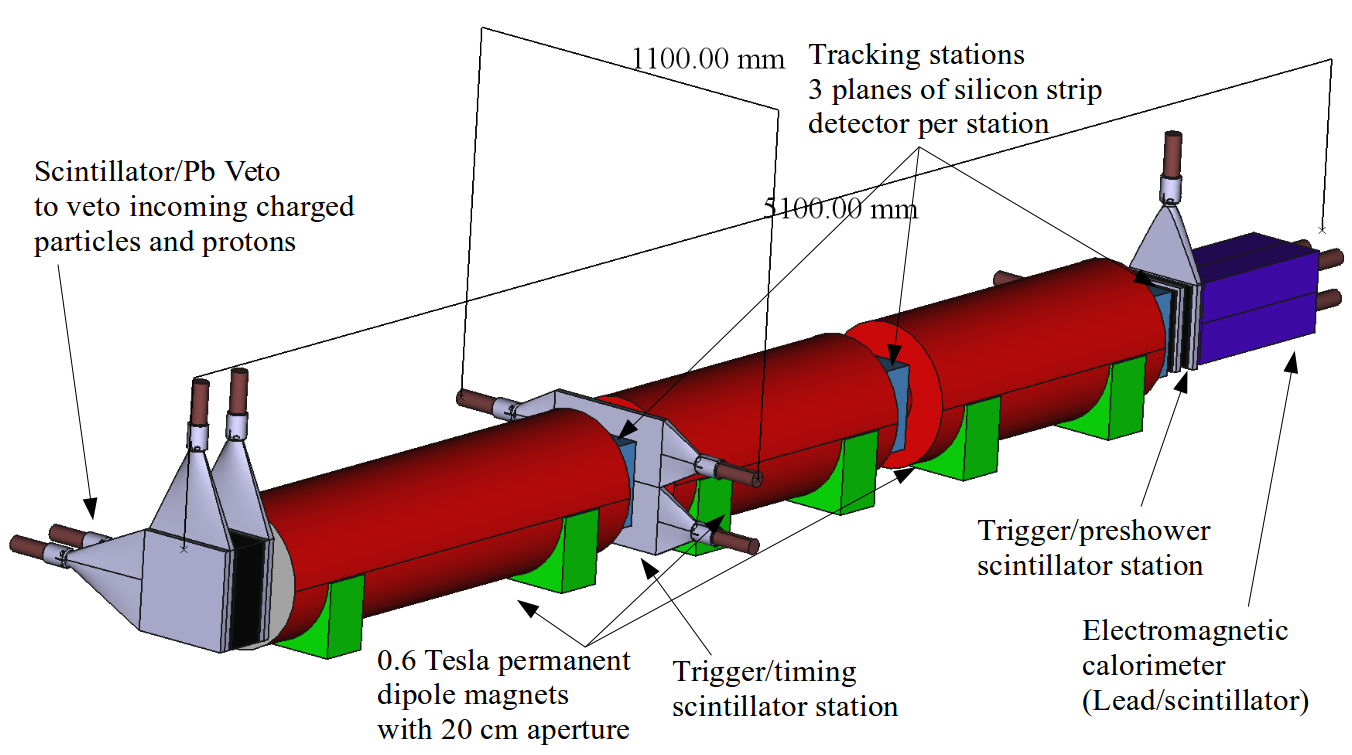} 
\caption{
Layout of the FASER detector. LLPs enter from the left and the entire length of the detector is roughly 5 m.  The detector components include scintillators (gray), dipole magnets (red), tracking stations (blue), a calorimeter (dark purple), and support structures (green). 
}
\label{fig:DetectorLayout}
\end{figure}

FASER has been designed to be sensitive to the many possible forms of light, weakly-interacting particles, and to differentiate signal from background.  FASER's detector layout and components are shown in \figref{DetectorLayout}.
  
Particles produced at the ATLAS IP enter the detector from the left.  At the entrance to the detector is a double layer of scintillators (gray) to veto charged particles coming through the cavern wall from the IP, primarily high-energy muons.  Between the scintillation layers is a 20-radiation-lengths-thick layer of lead that converts photons produced in the wall into electromagnetic showers that can be efficiently vetoed by the scintillators. By including two scintillator veto stations at the entrance to FASER, each able to detect at least $99.99\%$ of the incoming high-energy muons, the leading background of radiative processes associated with muons coming from the IP can be reduced to negligible levels. 

The veto layer is followed by an $L = 1.5~\text{m}$ long, 0.6 T permanent dipole magnet (red) with a $R= 10~\text{cm}$ aperture radius.  Such permanent magnets take up relatively little space and, unlike electromagnets, do not require high voltage power and cooling. The cylindrical volume enclosed by this magnet serves as the decay volume for the light, weakly-interacting particles, with the magnet providing a horizontal kick to separate oppositely-charged particles to a detectable distance. 

Next is a spectrometer consisting of two $1~\text{m}$-long, 0.6 T dipole magnets with three tracking stations (blue), each composed of layers of precision silicon strip detectors located at either end and in between the magnets. The primary purpose of the spectrometer is to observe the characteristic signal of two oppositely charged particles pointing back towards the IP, measure their momenta, and sweep out low-momentum charged particles before they reach the back of the spectrometer. Scintillator planes (gray) for triggering and precision time measurements are located at the entrance and exit of the spectrometer.  

The final component is an electromagnetic calorimeter (purple) to identify high energy electrons and photons and measure the total electromagnetic energy.  As the primary signals are two close-by electrons or photons, these cannot be resolved by the calorimeter.

\section{Discovery Potential}
\label{sec:discovery}

FASER's sensitivity to new physics has been evaluated in the context of many specific models~\cite{Feng:2017uoz, Feng:2017vli, Batell:2017kty, Kling:2018wct, Helo:2018qej, Bauer:2018onh, Cheng:2018vaj, Feng:2018noy, Hochberg:2018rjs, Berlin:2018jbm, Dercks:2018eua}.  In this section, we give a brief overview of some of these studies.    We present results for FASER as described above and designed to collect data during LHC Run 3 from 2021-23; and FASER 2, which may collect data in the HL-LHC era from 2026-35. For FASER 2, following the FASER design, we assume a cylindrical shape with depth $L$ and radius $R$.  The parameters for these two detectors and the assumed integrated luminosity for each of them, then, are 
\be
\textbf{FASER:} \ \  & L= 1.5~\m, &R& = 10~\cm, &\mathcal{L}& =150~\ifb  \\
\textbf{FASER 2:} \ \  & L = 5~\m,  &R& = 1~\m,     &\mathcal{L}& =3~\iab \ .
\label{eq:geomtery}
\ee
The collision energy is assumed to be 14 TeV in both cases, and, as with FASER, we assume FASER 2 will be located 480 m from the IP.  At present, the design of FASER 2 has not been carefully studied, and the FASER 2 parameters should only be taken as representative of a detector that is much larger than FASER.   In determining the physics reach for the various models below, we assume 100\% signal efficiency and negligible backgrounds. For FASER, these assumptions are based on simulation studies. It should also be noted that the sensitivity contours are only slightly affected by ${\cal O}(1)$ changes in the detection efficiency. 

\begin{figure}[tbp]
\centering
\vspace*{-0.6cm}
\includegraphics[trim={17.6cm 0 0 0}, clip, width=0.48\textwidth]{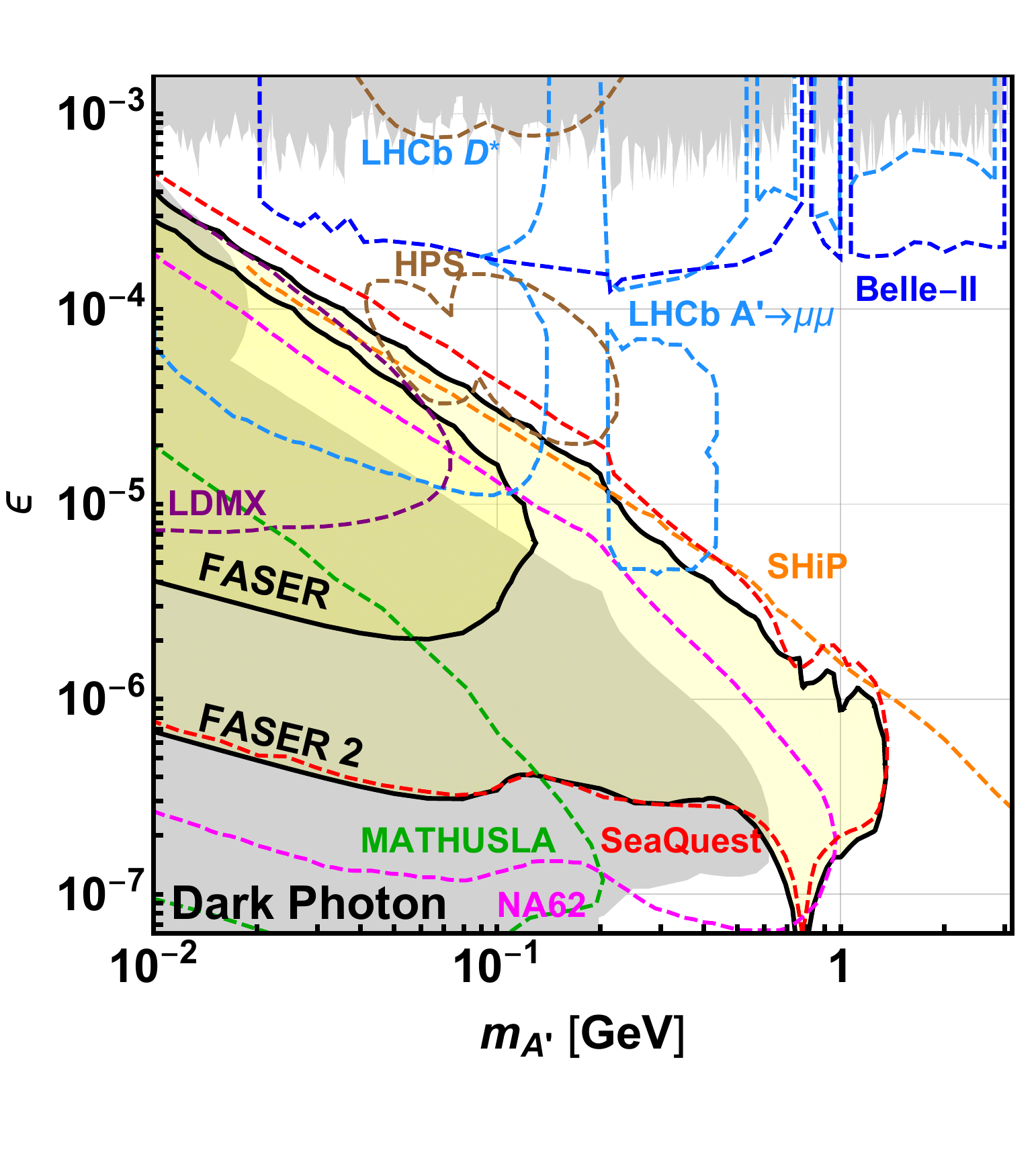} 
\hfill
\includegraphics[trim={17.6cm 0 0 0}, clip, width=0.48\textwidth]{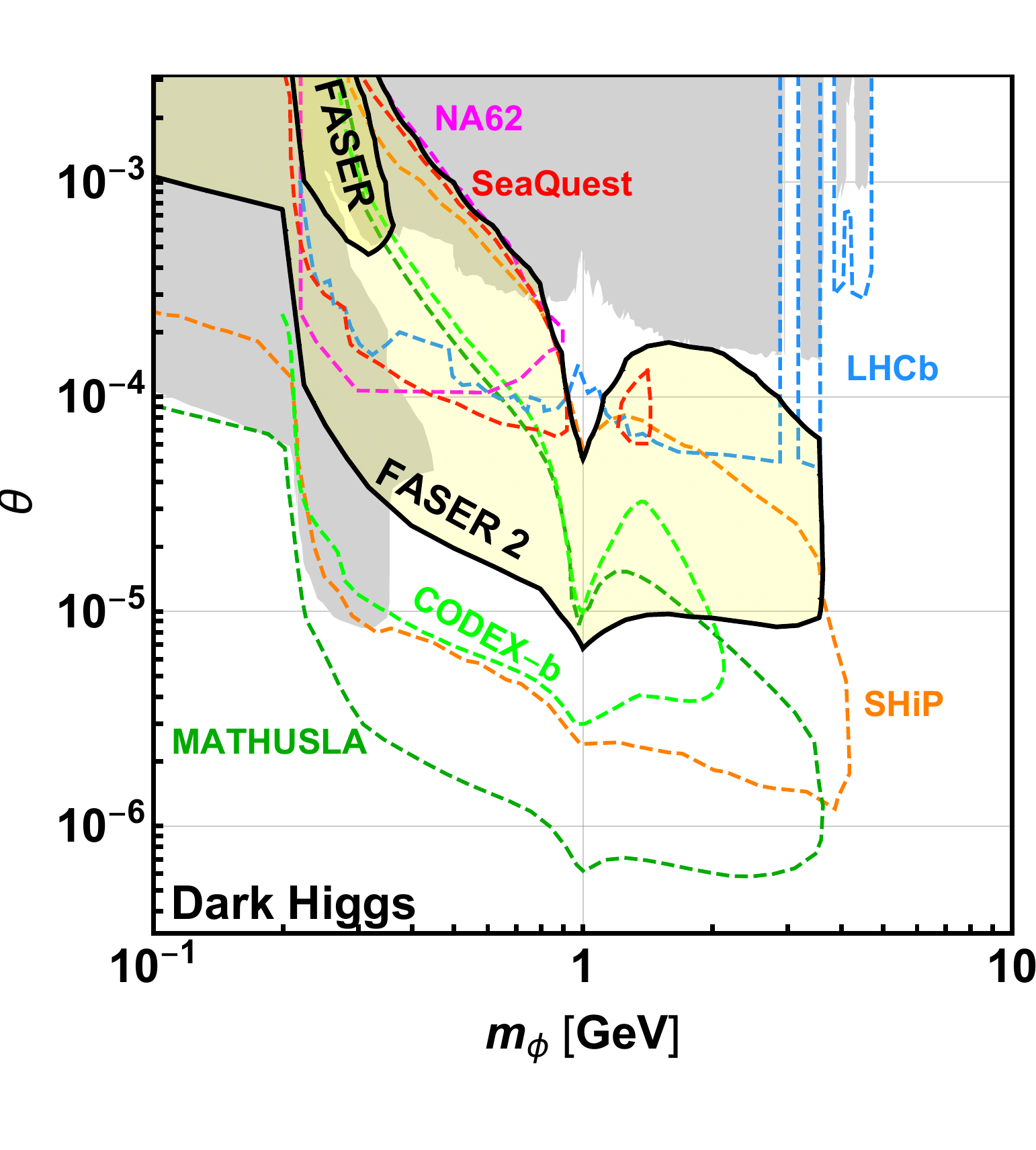} \\
\includegraphics[trim={17.6cm 0 0 0}, clip, width=0.48\textwidth]{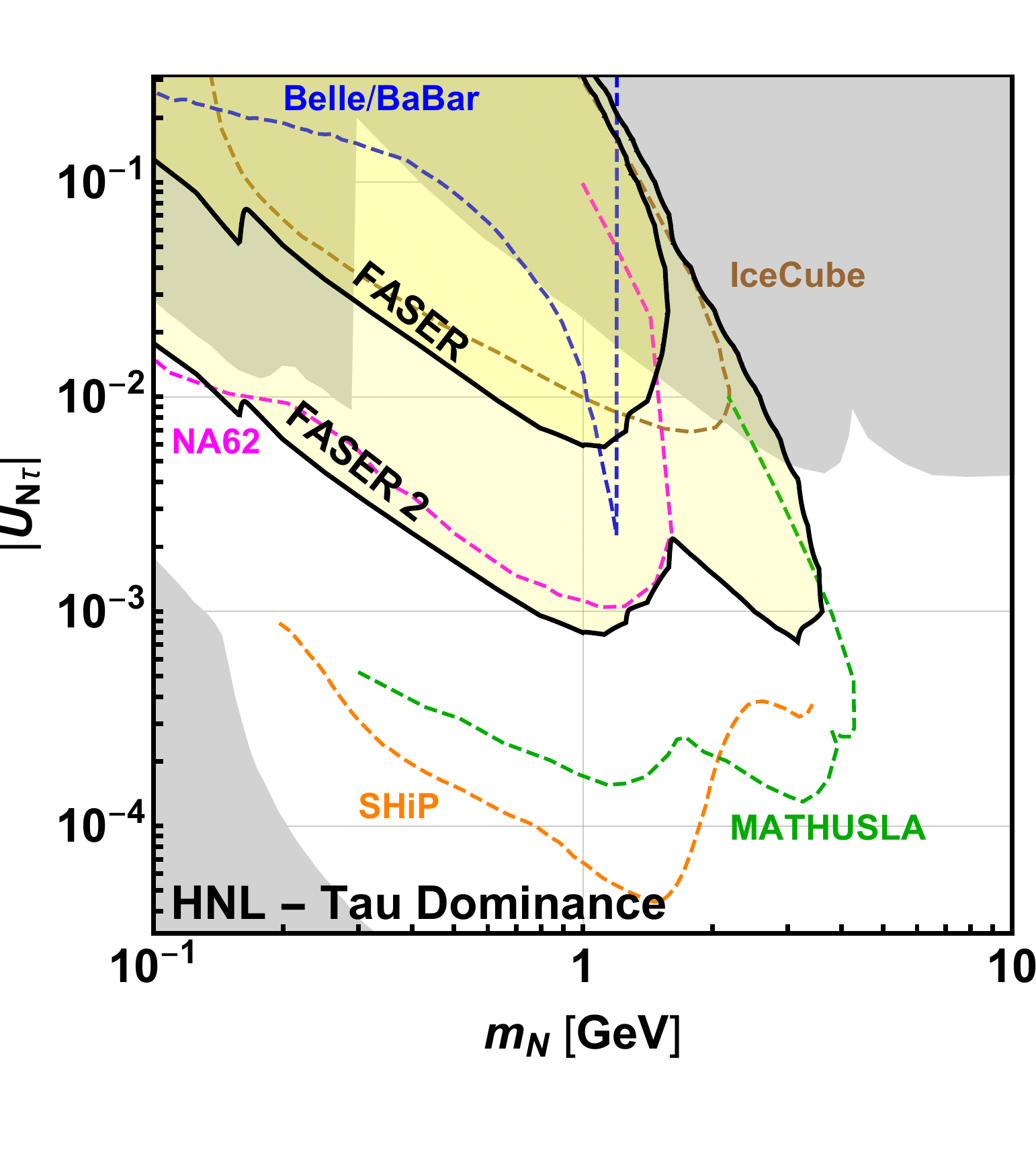} 
\hfill
\includegraphics[trim={17.6cm 0 0 0}, clip, width=0.48\textwidth]{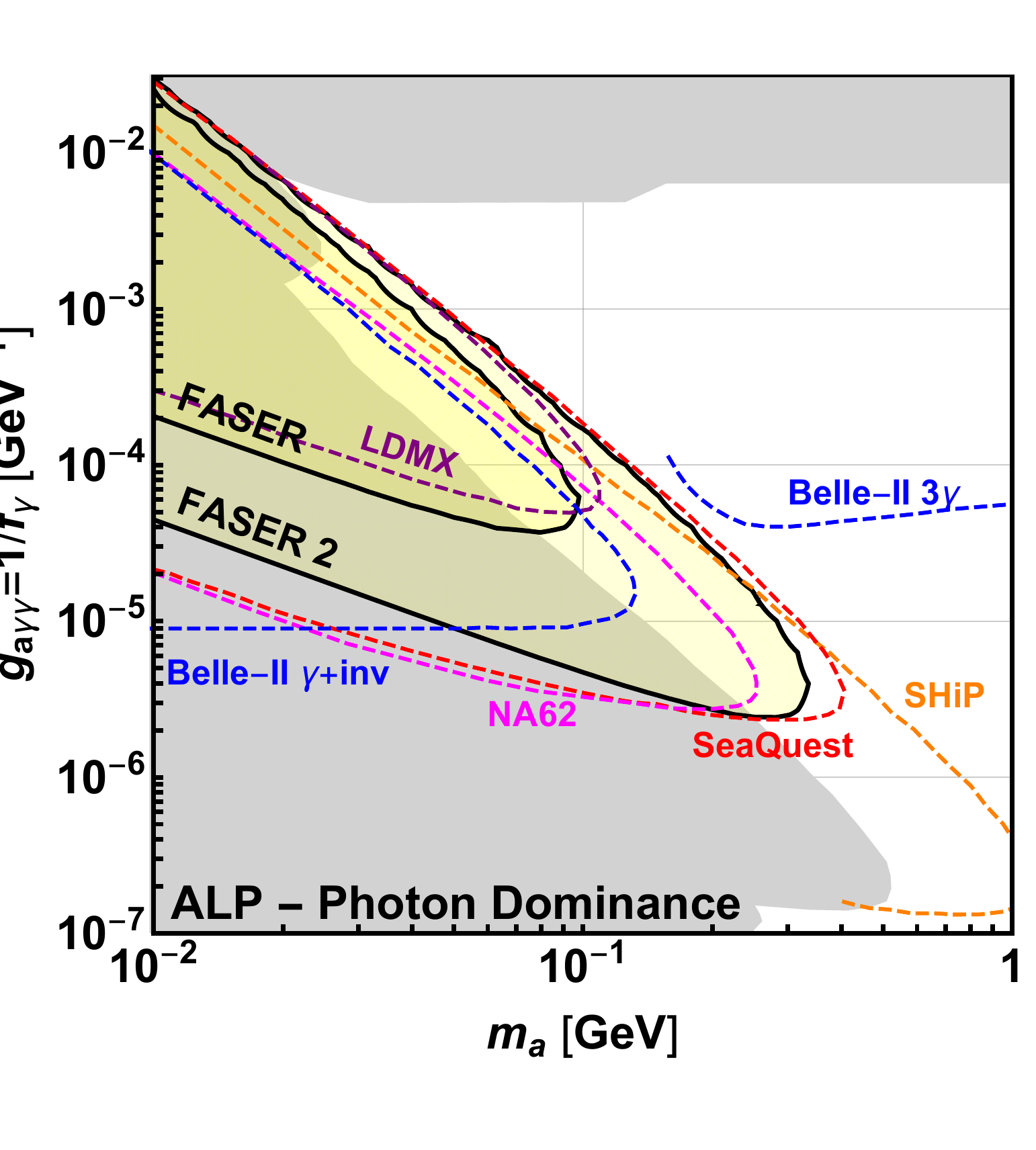} 

\caption{Sensitivity reaches for FASER and FASER 2 for (upper left) dark photons, (upper right) dark Higgs bosons, (lower left) heavy neutral leptons (HNLs) with dominantly $\tau$-mixing, and (lower right) axion-like particles (ALPs) with dominantly photon couplings.  The gray-shaded regions are excluded by current bounds, and the projected future sensitivities of other experiments are shown as colored contours.  See the text for details.}
\label{fig:discovery}
\end{figure}

The sensitivity reaches for 4 representative models for long-lived particles are shown in \figref{discovery}.  For dark photons, we see that FASER will probe new parameter space with masses $m \sim 100~\mev - \gev$ and couplings $\epsilon \sim 10^{-5} - 10^{-4}$.  FASER 2 will extend coverage to much larger masses $m \sim 1~\gev$, and to larger couplings $\epsilon$.  FASER 2's reach is comparable to SHiP at high $\epsilon$, although less sensitive at masses $\agt \gev$ and low $\epsilon$.  

The dark Higgs boson case is interesting, in that the dominant production mechanism is through heavy meson decay.  FASER therefore benefits from the large number of $B$ mesons produced at the LHC, as noted above, but at the same time, the long-lived particles produced are less collimated.  FASER has little reach in this model, but FASER 2, with its larger aperture, is able to probe significant new regions of parameter space through the process $b \to s \phi$, where $\phi$ is the dark Higgs boson.  Interestingly, FASER 2 can also probe the trilinear scalar coupling $h \phi \phi$ by looking for the process $b \to s h^* \to s \phi \phi$~\cite{Feng:2017vli}. In these ways, FASER 2 can search for new particles and also sensitively probe the properties of the standard model Higgs boson, one of the most important goals for experimental efforts in the coming decade.  Note the complementarity in this case between FASER and other proposed long-lived particle searches, including MATHUSLA, CODEX-b, and SHiP.

Finally, results are shown for heavy neutral leptons that dominantly mix with the $\tau$ multiplet and ALPs that are dominantly coupled to photons.  In both cases, FASER can discover new physics and FASER 2 significantly extends this reach into unprobed parameter space. 

Of course, FASER has sensitivity to many other proposed candidates for light, weakly-interacting new particles.  These include other benchmark models that are part of the Physics Beyond Colliders study~\cite{PBCwebpage}, as well as others, including inelastic dark matter~\cite{Berlin:2018jbm}, $R$-parity violating supersymmetry~\cite{Helo:2018qej,Dercks:2018eua}, models with strongly interacting massive particles (SIMPs)~\cite{Hochberg:2018rjs}, and twin Higgs scenarios~\cite{Cheng:2018vaj}.  A summary of FASER and FASER 2's discovery prospects for some of these models is given in \tableref{summary}.  

\begin{table}
  \centering
  \begin{tabular}{|c|c|c|c|c|c|c|}
  \hline \hline
	{\bf Benchmark Model} 
	& \, {\bf PBC} \ & \, {\bf Refs} \ & \, {\bf FASER} \ & \, {\bf FASER 2} \ \\ \hline \hline
    Dark Photons 
    & BC1 & \cite{Feng:2017uoz} & $\surd$ & $\surd$ \\ 
    $B-L$ Gauge Bosons 
    & ---  & \cite{Bauer:2018onh} & $\surd$ & $\surd$ \\ 
    $L_i - L_j$ Gauge Bosons 
    & --- & \cite{Bauer:2018onh} & --- & --- \\ \hline
    Dark Higgs Bosons 
    & BC4 & \cite{Feng:2017vli,Batell:2017kty} & --- & $\surd$ \\ 
    Dark Higgs Bosons with $hSS$ 
    & BC5 & \cite{Feng:2017vli} & --- & $\surd$ \\ \hline
    HNLs with $e$ 
    & BC6 & \cite{Kling:2018wct,Helo:2018qej} & --- & $\surd$ \\ 
    HNLs with $\mu$ 
    & BC7 & \cite{Kling:2018wct,Helo:2018qej} & --- & $\surd$ \\ 
    HNLs with $\tau$ 
    & BC8 & \cite{Kling:2018wct,Helo:2018qej} & $\surd$ & $\surd$ \\ \hline
    ALPs with Photon 
    & BC9 & \cite{Feng:2018noy} & $\surd$ & $\surd$ \\ 
    ALPs with Fermion 
    & BC10 & \cite{Ariga:2018uku} & $\surd$ & $\surd$ \\ 
    ALPs with Gluon 
    & BC11 & \cite{Ariga:2018uku} & $\surd$ & $\surd$ \\ 
    \hline \hline
  \end{tabular}
  \caption{FASER's discovery prospects in a variety of benchmark models.  For each model, we list their PBC labels, references in which they were previously studied, and the prospects for FASER and FASER 2 to probe new parameter space.  FASER and FASER 2 have discovery potential for all candidates with renormalizable couplings (dark photons, dark Higgs bosons, HNLs); ALPs with all types of couplings ($\gamma$, $f$, $g$); and also other models not listed here~\cite{Cheng:2018vaj, Hochberg:2018rjs, Berlin:2018jbm,Dercks:2018eua}.} 
\label{table:summary}
\end{table}

In addition to searches for new physics, FASER may also provide interesting probes of standard model physics.  The FASER program benefits from the significant progress that has been made in recent years in understanding high-energy forward physics~\cite{N.Cartiglia:2015gve}, and FASER will, in turn, add to our understanding of both forward physics and cosmic ray physics.  In addition, the FASER program has strong prospects for providing new insights into neutrinos. As an example, in Run 3, the number of muon neutrinos passing through FASER with energies above 100 GeV is $\sim 10^{13}$, with roughly 600 interacting in the 10 cm-thick block of lead that is near the front of FASER.  Furthermore, a few tau neutrinos with $\sim \tev$ energies are expected to interact in FASER. Although more study is required, these event rates imply that FASER, and especially FASER 2, may also provide interesting information about SM particles by detecting the first neutrinos at the LHC and, for example, constraining neutrino interaction rates in the energy range $E_{\nu} \sim 400~\gev - 4~\tev$, where they are currently unconstrained.

\section{Current Status and Anticipated Challenges}
\label{sec:summary}

The null results of new physics searches in the high-$p_T$ region of $pp$ collisions call for new ideas that could extend the LHC physics reach. The FASER program will extend the LHC's physics program by searching for new light, weakly coupled LLPs in the far forward region of $pp$ collisions, with the potential to discover physics beyond the SM and shed light on dark matter. 

In its first stage, FASER is an extremely small and inexpensive detector with a decay volume of only $0.047~\m^3$.  It is at an advanced stage in obtaining CERN approval and two private foundations have expressed their intention to provide roughly 2 MCHF in funding, sufficient to support FASER's construction and some operation costs.  If approved, the detector will be installed in TI12 during LS2 and take data during Run 3, probing new regions of parameter space for dark photons, other light gauge bosons, HNLs with dominantly $\tau$ couplings, and axion-like particles with masses in the 10 MeV to GeV range. FASER will run concurrently with the other LHC experiments, requiring no beam modifications and interacting with the accelerator and existing experiments only in requesting luminosity information from ATLAS and bunch crossing timing information from the LHC.    

A larger detector, FASER 2, running in the HL-LHC era, could extend this sensitivity to larger masses and will probe currently unconstrained parameter space for all renormalizable portals (dark photons, dark Higgs bosons, and heavy neutral leptons), ALPs with photon, fermion, or gluon couplings, and many other new particles.  It is important to note, however, that the FASER 2 detector, with dimensions considered here of $R \sim 1~\m$ and $L \sim 5~\m$, will require significant excavation to extend either TI12 or TI18, or to widen the cavern UJ18 near TI18 or the cavern UJ12 near TI12.  In addition, the cost of a magnet surrounding such a large decay volume may well be prohibitive.  As FASER 2 becomes more studied, it will be important to develop new designs that do not simply scale up FASER, but nevertheless preserve the relatively low cost of FASER, while optimizing FASER 2's discovery prospects.

\appendix

\section*{Appendix: Addendum}
\label{sec:addendum}

This Addendum to the main FASER document describes the community interested in FASER, as well as FASER's timeline, construction and operating costs, and computing requirements. 

\subsection{Interested Community}

The FASER Collaboration currently includes the following people:
\begin{itemize}
\item Experimentalists: Jamie Boyd (CERN, co-spokesperson), Akitaki Ariga (Bern), Tomoko Ariga (Bern and Kyushu), Franck Cadoux (Geneva), David Casper (UC Irvine), Yannick Favre (Geneva), Didier Ferrere (Geneva), Shih-Chieh Hsu, (Washington), Giuseppe Iacobucci (Geneva), Enrique Kajomovitz (Technion), Susanne Kuehn (CERN), Lorne Levinson (Weizmann), Hidetoshi Otono (Kyushu), Brian Petersen (CERN), Osamu Sato (Nagoya), Matthias Schott (Mainz), Sergio Sevilla (Geneva), Anna Sfyrla (Geneva), Aaron Soffa (UC Irvine), Yosuke Takubo (KEK), Eric Torrence (Oregon), Gang Zhang (Tsinghua, China).
\item Theorists: Jonathan Feng (UC Irvine, co-spokesperson), Iftah Galon (Rutgers), Felix Kling (UC Irvine), Jordan Smolinsky (UC Irvine), Sebastian Trojanowski (National Centre for Nuclear Research, Warsaw and Sheffield).
\end{itemize}
In addition, a large number of CERN staff, although not members of the Collaboration, are supporting FASER in essential ways.  Many of them and their contributions are listed in FASER's Letter of Intent~\cite{Ariga:2018zuc} and Technical Proposal~\cite{Ariga:2018pin}.

For FASER 2, the Collaboration can be expected to grow considerably, given the large and growing interest in light and weakly-interacting particles.  The broader community includes collaborators on complementary experiments with similar physics targets, including HPS~\cite{Moreno:2013mja}, Belle-II~\cite{Dolan:2017osp}, LHCb~\cite{Ilten:2015hya, Ilten:2016tkc}, NA62~\cite{Dobrich:2018ezn}, NA64~\cite{Gninenko:2018tlp}, SeaQuest~\cite{Berlin:2018pwi}, SHiP~\cite{Alekhin:2015byh}, MATHUSLA~\cite{Evans:2017lvd,Curtin:2018mvb}, CODEX-b~\cite{Gligorov:2017nwh}, AL3X~\cite{Gligorov:2018vkc}, and LDMX~\cite{Berlin:2018bsc}.  For a more complete list of current and proposed experiments, see, e.g., Ref.~\cite{Battaglieri:2017aum}.

Of course, in addition to those who might work on FASER, the community interested in FASER includes all who are interested in the fields of physics FASER will impact.  These fields include physics beyond the standard model, dark matter, particle cosmology and astrophysics, neutrino physics, flavor physics, and also possibly cosmic ray physics.

\subsection{Timeline}

The timeline for the FASER program is set by the LHC running schedule.  The FASER detector will be constructed in Long Shutdown 2 from 2019-20 in time to collect data during the LHC's Run 3 from 2021-23.  After successful operation of FASER, FASER 2 could be constructed and installed during LS3 from 2024-26 and collect data during the HL-LHC era from 2026-35.

\subsection{Construction and Operating Costs}

An essential feature of the FASER program is its ability to do world-leading physics at a very affordable cost, thanks to the size and location of the experiments.  FASER's decay volume is just $0.047~\m^3$, and the entire experiment fits in a box with dimensions $1~\m \times 1~\m \times 5~\m$.  In addition, FASER's location in TI12 is quiet, so detector components do not need to be radiation hard, and background radiation for electronics is not a great concern.   

A detailed cost estimate of FASER has been presented in the FASER Technical Proposal~\cite{Ariga:2018pin}.  The total cost for construction is approximately 800 kCHF, with roughly half of the cost in magnet construction. FASER's low cost and rapid construction schedule are made possible by re-use of hardware from other experiments. In particular, the ATLAS silicon tracker (SCT) and LHCb Collaborations have generously allowed FASER to use their spare tracker and calorimeter modules, respectively.    The entire cost of FASER construction and installation, as well as some operating expenses, is expected to be funded by two private foundations through grants totalling roughly 2 MCHF.  An additional cost of roughly 300 kCHF, largely civil engineering and transport costs, is expected to be borne by CERN. Planning for FASER 2 is still in its early stages, but its cost may be roughly $\sim 5-10$ times that of FASER.

\subsection{Computing Requirements}

FASER's computing and storage requirements will be orders of magnitude smaller than the flagship LHC experiments. The detector's 72 ATLAS SCT modules represent less than 2\% of the corresponding ATLAS sub-detector, and the occupancy is lower by a similar factor.  Reconstruction of FASER data will consume negligible CPU time. Simulation time is dominated by showers in the calorimeter, and can be made negligible by standard parameterization and fast simulation techniques.

Work on FASER 2 is still very preliminary, but as with the construction cost, its computing requirements would be somewhat greater than FASER's. Depending on the detector technology, the occupancy per event may not be much different, but the trigger rate will scale with the detector's larger area and the HL-LHC's higher luminosity. Even so, these requirements will remain insignificant compared to those of the large LHC experiments.

\bibliography{FASER_ESPP}

\end{document}